\documentclass[12pt,preprint,preprintnumbers,nofootinbib,
groupedaddress,superscriptaddress,amsmath,amssymb]{revtex4}

\usepackage{amsmath,amssymb}
\usepackage{slashed}
\usepackage{graphics}
\usepackage[dvipdfm]{graphicx}

\usepackage{cancel,graphics,subfigure}

\usepackage{color}

\begin{document}
\title{Scalar leptoquarks and Higgs pair production at the LHC}
\author{Tsedenbaljir Enkhbat}
\email{enkhbat@gmail.com}
\affiliation{
Institute of Physics \& Technology, Mongolian Academy of Sciences, \\
Ulaanbaatar 13330, Mongolia \&\\
Department of Physics and Center for Theoretical Sciences,
National Taiwan University, Taipei 10617, Taiwan}



\begin{abstract}
\noindent
The presence of colored particles can affect both the single and the pair Higgs productions substantially. For scalar particles, this happens if their portal couplings to the Standard Model Higgs are large and their masses are not too high. In the present work  these processes are studied in the case of several leptoquarks which may appear in many beyond Standard Model theories. 
It is found that the constraints on the portal couplings  from the single Higgs production and the decays to various channels measured by the LHC experiments still allow increased Higgs pair production rate. For the masses in the range from 180~GeV to 300~GeV, depending on the strength of such portal couplings,  the Higgs pair production
may reach an order to several hundred in magnitude larger rate than the Standard Model case for the 8~TeV run. Therefore, combined with the on going searches for leptoquarks by both the CMS and ATLAS, this is one of the possible scenarios to be probed directly by the current data. The current study demonstrates that if colored scalars modify scalar potentials through portal couplings, which has been studied for variety of motivations such as playing a potentially important role in electroweak phase transition, composite models or radiative neutrino masses, this fact may appear as the modified Higgs pair production. 
\end{abstract}

\maketitle


\section{Introduction}

\noindent The data collected by the LHC experiments at 7 and 8~TeV with $\sim$5 and 20fb$^{-1}$ respectively is refining the 
details of the Higgs like resonances found last year~\cite{atlas:2012gk, cms:2012gu}. Many decay channels have been searched for and the individual channels 
so far have given us a consistent picture with what one expects 
from the SM Higgs. On the other hand, the self interaction of the 
Higgs, which is probed by the Higgs pair production ~\cite{Eboli:1987dy, Glover:1987nx, Dicus:1987ic, Plehn:1996wb, Dawson:1998py},   is too feeble in the SM to be detected with these early data set. Even at 14~TeV run, the luminosity  required for 
probing this process is very high~\cite{Dawson:1998py, Djouadi:1999rca, Dittmaier:2011ti, Branco:2011iw, Baglio:2012np, Dolan:2012rv, Papaefstathiou:2012qe, Goertz:2013kp, deFlorian:2013uza, Grigo:2013rya, Barger:2013jfa}. This fact, namely the smallness of the corresponding Higgs pair production cross--section, makes it prone to a presence of a new physics~\cite{Belyaev:1999mx, Asakawa:2010xj, Kribs:2012kz, Dorsner:2012pp, Heng:2013wia, Wang:2007zx, Han:2009zp, Dolan:2012ac, Han:2013sga, Hao:2012ir, Wang:2012zv, Shao:2013bz, No:2013wsa, Haba:2013xla}.

In particular, relatively light colored particles are known to affect the cross-section substantially~\cite{Belyaev:1999mx, Asakawa:2010xj, Kribs:2012kz, Dorsner:2012pp, Heng:2013wia}. As a mater of fact there are many models with various motivations  including models of GUT remnants~\cite{Georgi:1974sy, Pati:1974yy, Senjanovic:1982ex, Shanker:1982nd, Buchmuller:1986iq, Buchmuller:1986zs, Angelopoulos:1986uq, Hewett:1988xc}, 
composite models~\cite{Dimopoulos:1979es, Dimopoulos:1979sp, Eichten:1979ah, Farhi:1980xs, Schrempp:1984nj, Wudka:1985ef, Gripaios:2009dq, Hung:2009hy, Enkhbat:2011vp} or in a radiative neutrino mass models~\cite{Babu:2010vp, Babu:2011vb, Kohda:2012sr} which may give such contributions. Among these the scalars are interesting as they may play crucial role in the spontaneous symmetry breaking through additional terms with large portal couplings in the scalar potential.  Furthermore, another reason to be interested in colored scalars is that they are known to have a potentially crucial role on achieving a successful electroweak phase transition (EWPT). Common feature of these models is that the colored particle(s) must be light enough for a strong enough  EWPT~\cite{Pietroni:1992in, Cline:1998hy, Cline:1999wi, Carena:2008vj, Cohen:2011ap, Chung:2012vg, Huang:2012wn, Laine:2012jy, Patel:2013zla}. The discovery of  the new resonance has triggered renewed interest in colored particles from this point of view  and several groups have made detailed studies. Multiple scalars tend to broaden available parameter space for EWPT. For example, the so called light stop scenario has been the subject of a recent study~\cite{Huang:2012wn, Patel:2013zla}. Due to their possible importance it is crucial to study more broader class of models with colored scalars.

In the present work we study the phenomenological consequences of the Standard Model extension by two or more colored scalar particles.  As a case study we take several leptoquarks (LQ) since there is an active experimental program by both ATLAS and CMS for the search~\cite{Chatrchyan:2012vza, cmsLQ2, Chatrchyan:2012st, Chatrchyan:2012sv, ATLAS:2013oea, Stupak:2012aj}. The LHC search for an individual LQ have  now reached as high as 830~GeV, 525~GeV with 5~fb$^{-1}$ 7~TeV data  for first and third generation LQs, 1070~GeV with 20~fb$^{-1}$ 8~TeV data for second generation LQ respectively 
 assuming they decay 100\% to the considered decay channels. If the LQ masses are above these limits, their effect on the Higgs 
 phenomenology would be very minimal.
On the other hand simultaneous presence of several LQs, may open up additional channels  and therefore weakens these bounds. Specific 
models where the LQs are introduced to explain a certain phenomenon usually requires more than one LQs as in the model we study here. 

I examine a possibility of the existence of LQs with masses as light as $\sim$200~GeV and study their effect for the single and di Higgs productions. As we will see the Higgs pair production is substantially altered in the low mass range below 300~GeV without too much change in the Higss  diphoton decay channel if portal couplings are large. These couplings are required to have opposite signs by the latest Higgs data  or small in magnitude.
The model I consider has two LQs, an $SU(2)$ doublet $\omega$ and a singlet $\chi$. As we will see their simultaneous presence still allows them to have relatively light masses and escape the current bounds.  In particular, the current bounds do not include  LQs decaying to $\mu t$ or $\tau t$. Such a scenario, for example, has appeared in  a model considered by Babu and Julio \cite{Babu:2010vp}, where the light neutrino masses are induced by two--loop effects from LQs.
If their masses are only of order few hundred GeV, as it is required in this case, the scenario can be  probed or even excluded with the data from the LHC. Therefore this is one of the easiest model which can be tested and is the subject of the current study. Although I consider a particular model, it should be stressed that other models with colored particles can affect the pair productions in a similar manner.

In Section~II, I briefly list the current experimental status on the Higgs production and decay rates. Then I introduce the model I examined in the paper. Section~III contains main part of this work where the numerical results for the single and pair Higgs productions are presented. 
The conclusion is given in Section~IV.



\section{Light LeptoQuarks}
\noindent
ATLAS and CMS both have released their results on the Higgs searches from 7 and 8~TeV runs. The median significance of the diphoton channel for ATLAS, while remains above the SM level, has come down to $\mu_{\gamma\gamma}=1.53\substack{+0.34 \\ -0.3}$~\cite{atlasdiboson} compared to the 7~TeV result. On the other hand the change in the latest CMS result compared to its 7~TeV data was more dramatic. Depending on the analysis the signal strength  now stands either at $\mu_{\gamma\gamma}=0.78\substack{+0.28 \\ -0.26}$ or $1.11\substack{+0.32 \\ -0.30}$~\cite{cmsdiphot}. Also importantly, the measurements for $h\rightarrow ZZ^*\rightarrow 4\ell$ channel strength are $\mu_{4\ell}=0.91\substack{+0.30\\-0.24}$ from CMS~\cite{cmszz} and $\mu_{4\ell}=1.7\substack{+0.5\\-0.4}$~\cite{atlasdiboson} from  ATLAS respectively which constrain the production separately. These results indicate that the diphoton channel of Higgs decay is closer to the SM prediction than it has  appeared from the 7~TeV data.  
Therefore, any new resonance should not affect the single Higgs production and the diphoton channel too much. This requirement alone makes a single colored scalar object harder to exist at lower mass range if its portal coupling of $|H|^2|X|^2$ type is large. If such couplings are small they will not play any interesting role in the Higgs phenomenology. On the other hand several colored scalars can lead to interesting excesses that may be checked with the existing data at the same time satisfying various Higgs decay channels measurements. 

The model I examine in this paper  contains two new multiplets, $SU(2)_L$ singlet and doublet scalar leptoquarks $\Omega\sim(3,2,1/6)$ and $\chi\sim(3,1,-1/3)$~\cite{Babu:2010vp}. 
 The Lagrangian of the model is given as:
\begin{eqnarray}\label{eq:lag}
{\cal L}&=&\left(Y_{ij}\Omega i \sigma_2 L_i d^c_j+F_{ij}\chi e^c_iu^c_j-\mu \Omega^\dagger H \chi+\text{h.c}\right)-m_\Omega^2 |\Omega|^2-m_\chi^2|\chi|^2\nonumber\\ &-&\lambda_{\omega}|\Omega|^2 |H|^2-\lambda_{\chi}|\chi|^2 |H|^2-\kappa|\Omega^\dagger H|^2
\end{eqnarray}
After electroweak symmetry breaking, the lower component of the doublet LQ will mix with the singlet LQ via the trilinear $\mu$--term which we denote as $\chi_1$ and  $\chi_2$, and the remaining upper $2/3$ charged component as $\omega$. Their physical masses are given by
\begin{eqnarray}\label{eq:mass1}
m_\omega^2&=&m_\Omega^2+\frac{\lambda_\omega}{2} v^2,\label{eq:momega}\\
m_{\chi_1,\chi_2}^2&=&\frac{1}{2}\left(m_\omega^2 + \frac{\kappa}{2} v^2+m_\chi^2 + \frac{\lambda_\chi}{2} v^2 \mp\sqrt{m_\omega^2 + \frac{\kappa}{2} v^2-m_\chi^2- \frac{\lambda_\chi}{2} v^2+2\mu^2 v^2}\right),\label{eq:mchi}\\
\tan2\vartheta&=&\frac{2\sqrt{2}\mu v}{2m_\omega^2 + \kappa v^2-2m_\chi^2-\lambda_\chi v^2},\label{eq:theta}
\end{eqnarray}
where $\vartheta$ and $m_{\chi_1,\chi_2}$ are the mixing angle and masses for the $-1/3$ charged LQs $\chi_1$ and $\chi_2$. $m_\omega$ is the mass of $2/3$ charged component denoted as $\omega$.
This spectrum was proposed by Babu and Julio as an explanation for the light  neutrino masses induced by two--loop effects of the LQs.  Readers interested in are referred to the original paper where exhaustive list of many flavor implications were discussed. Several scenarios in the model requires these LQs to be lighter than 500~GeV, which makes them testable at the LHC. I concentrate  primarily on the portal couplings and study their collider aspect and examine the consequences.

The searches for LQs at LHC have given lower bounds on their masses for several different LQ decay channels for the data collected at 7~TeV by both CMS and ATLAS collaborations. Assuming 100~\% branching fraction to electron or muon with a light quark, the pair produced LQs decaying to two leptons of the same flavor with at least two jets or single lepton with missing transverse energy and two jets have been ruled out up to 830~\cite{Chatrchyan:2012vza} with 7~TeV data and 1070~GeV~\cite{cmsLQ2} with full data for electron and muon channels respectively at 95\% confidence level by CMS collaboration. If the branching fractions are assumed to be 50~\% the limits are 630 and 840~GeV respectively. The third generation LQs are ruled out up to 450~GeV for $\nu\bar{b}$ by CMS~\cite{Chatrchyan:2012st}, and 525 and 535~GeV for $b\tau$  by CMS\cite{Chatrchyan:2012sv} and ATLAS~\cite{ATLAS:2013oea} respectively. For the $b\tau$ channel the bound from CMS weakens to $\sim$230~GeV if the branching fraction is $\sim$60\%. The ATLAS collaboration has not updated their searches for lighter generation LQs~\cite{Stupak:2012aj} beyond 7~TeV 1~fb$^{-1}$ data set. A thorough collider search analysis is beyond scope of this paper. Interested readers are referred to Refs~\cite{Kramer:2004df, Belyaev:2005ew, Gripaios:2009dq, Gripaios:2010hv, Davidson:2011zn}. In spite of all the above experimental advances in various channels, the searches for LQs decaying to $\mu t$ or $\tau t$ have not been done.  

If one considers any of the LQs, the LHC searches require that their masses have to be above 450~GeV. Unless corresponding portal 
coupling is very large the both single and di Higgs productions will not be affected at any interesting level. In the following we explain that these 
constraints may not be applicable for the model given by the Lagrangian in Eq.~(\ref{eq:lag}), To do so we consider a case where 
the following  mass hierarchy holds: $m_\omega>m_{\chi_2}>m_{\chi_1}$.  If the 
couplings $Y_{ij}$ in Eq~(\ref{eq:lag}) are small enough such that the mass splitting between $2/3$ and $-1/3$ charged LQ makes $\omega\rightarrow 
\chi_i W^{+*}\rightarrow \chi_i \bar f_d f_u$ channel dominant, these bounds are evaded. We call these three-body channels. Here $\bar f_d f_u= (\bar d u, \bar s c, \bar\ell \nu)$. The star signifies that the $W$ is off--shell. This is because the electroweak precision test requires the mass splitting within the $SU(2)_L$ doublet components be less than $\sim52$~GeV~\cite{pdg}.

Both the ATLAS and CMS have put the constraint on the mass by varying the branching fractions of the searched channels. ATLAS puts $\sim$10\% and 5\% upper bounds on the branching fraction for LQ decaying to $\mu q$ only and decaying equally to $\mu q$ and $\nu q$ respectively for LQ mass 200~GeV. The CMS has similar results but only down to 250~GeV Nevertheless, with larger data set of $5$~fb$^{-1}$, the branching fractions to $eejj$, $e\nu jj$, $ \mu\mu jj$ and $\mu\nu jj$ for LQ pair are constrained to be below $\sim12$, $2$, $12$ and $2$ percents respectively. The new data set from $8$~TeV will surely strenthen these further. In the scenario we consider the most stringent constraint comes from $\omega^{2/3}\rightarrow \ell q$ searches.  As for the $b\tau$ and $b\nu_\tau$ channels the branching fractions have to be below $\sim26$~\% and $\sim60$~\% respectively. To avoid these we assume the corresponding branching fraction be less than $\sim$10\% that of the three body channel. As long as we choose values small enough for $Y_{ij}$ satisfying the above inequality, the constraints from the searches for the light flavor LQs are avoided. Among the $F_{ij}$ couplings only $F_{23}$ and $F_{33}$ are allowed to be large by the LQ searches since they lead to the not--yet--searched $t\mu$ and $t\tau$ decay channels. Therefore we further assume the other $F_{ij}$ couplings are small and satisfy the constraints from various flavor changing neutral current constraints~\cite{Davidson:1993qk, Babu:2010vp, Babu:2011vb, Davidson:2010uu}. Further, if $F_{23}$ 
and/or $F_{33}$ are the largest $F_{ij}$ couplings, $\chi\rightarrow \mu\bar t $ or to $\tau \bar t$ will be the dominant $\chi_i$ decay  channel. The experimental bound $\tau\rightarrow \mu\gamma<4.4\times10^{-8}$ puts constraint $|F_{23}F_{33}^*|\lesssim0.2\times (m_{1}/200\text{GeV})^2$ which allows even a value of order one for either of these couplings.
From the above discussion we see that the signals for the  $\omega$ pair production are $\chi_i \chi_j W^{+*}W^{-*}$ with the off shell $W$'s subsequently decaying either hadronically or leptonically when $F_{23}$  or $F_{33}$ is the largest coupling.


\section{Higgs phenomenology with light LeptoQuarks}
\noindent
In the last section we have argued that the current limits from the LHC experiments still allow LQs with light masses down to $\sim$180~GeV.  Given that such a possibility exists in the current section we study their phenomenological consequences. The possibility that the Standard Model Higgs could have  portal couplings to an unknown sector has been a subject of many studies due to its possible role in the electroweak symmetry breaking, electroweak phase transition and as the contact with the dark sector. Recent discovery of the SM Higgs like resonance has intensified such studies.

The effect we investigate here is the Higgs boson pair production. We take several LQs and choose large portal couplings to demonstrate the di--Higgs production rate can be dramatically increased while the single Higgs production and diphoton rates are affected within the experimentally measured values. This will happen even with the current data if the LQs are relatively light below 300~GeV, which makes the model testable in most of the considered mass range. 

From the Lagrangian given in Eq.~(\ref{eq:lag}), the LQ and Higgs interactions are easily written down in the mass eigenstates as follows
\begin{eqnarray}\label{eq:portal}
V_{\text{\text{LQ}-h}}&=&\left\{\left(\lambda_\omega c^2_\vartheta+\kappa c^2_\vartheta+\lambda_\chi s^2_\vartheta\right)|\chi_1|^2+\left(\lambda_\omega s^2_\vartheta+\kappa s^2_\vartheta+\lambda_\chi c^2_\vartheta\right)|\chi_2|^2\right.\nonumber\\ &+& \left. \lambda_\omega |\omega|^2+\left(\lambda_\omega+\kappa-\lambda_\chi\right)s_\vartheta c_\vartheta\left(\chi_1\chi_2^*+\chi_2\chi_1^*\right)\right\}\left(\frac{h^2}{2}+hv\right)\nonumber\\ &+&\left\{\mu((|\chi_2|^2-|\chi_1|^2)c_\vartheta s_\vartheta+\chi_1^*\chi_2c^2_\vartheta-\chi_2^*\chi_1s^2_\vartheta)+\text{h.c}\right\}\frac{h}{\sqrt{2}},
\end{eqnarray}
where $s_\vartheta\,(c_\vartheta)\equiv\sin\vartheta\,(\cos\vartheta)$. We choose the physical masses $m_{\omega,\chi_{(1,2)}}$, portal couplings $\lambda_\omega$, $\lambda_{\chi}$ and the mixing angle as  the input parameters. Then the remaining parameters $\mu$ and $\kappa$ are fixed through Eqs.~(\ref{eq:momega}--\ref{eq:theta}).

\begin{figure}[!ht]
 \centering
 \includegraphics[height=16cm]{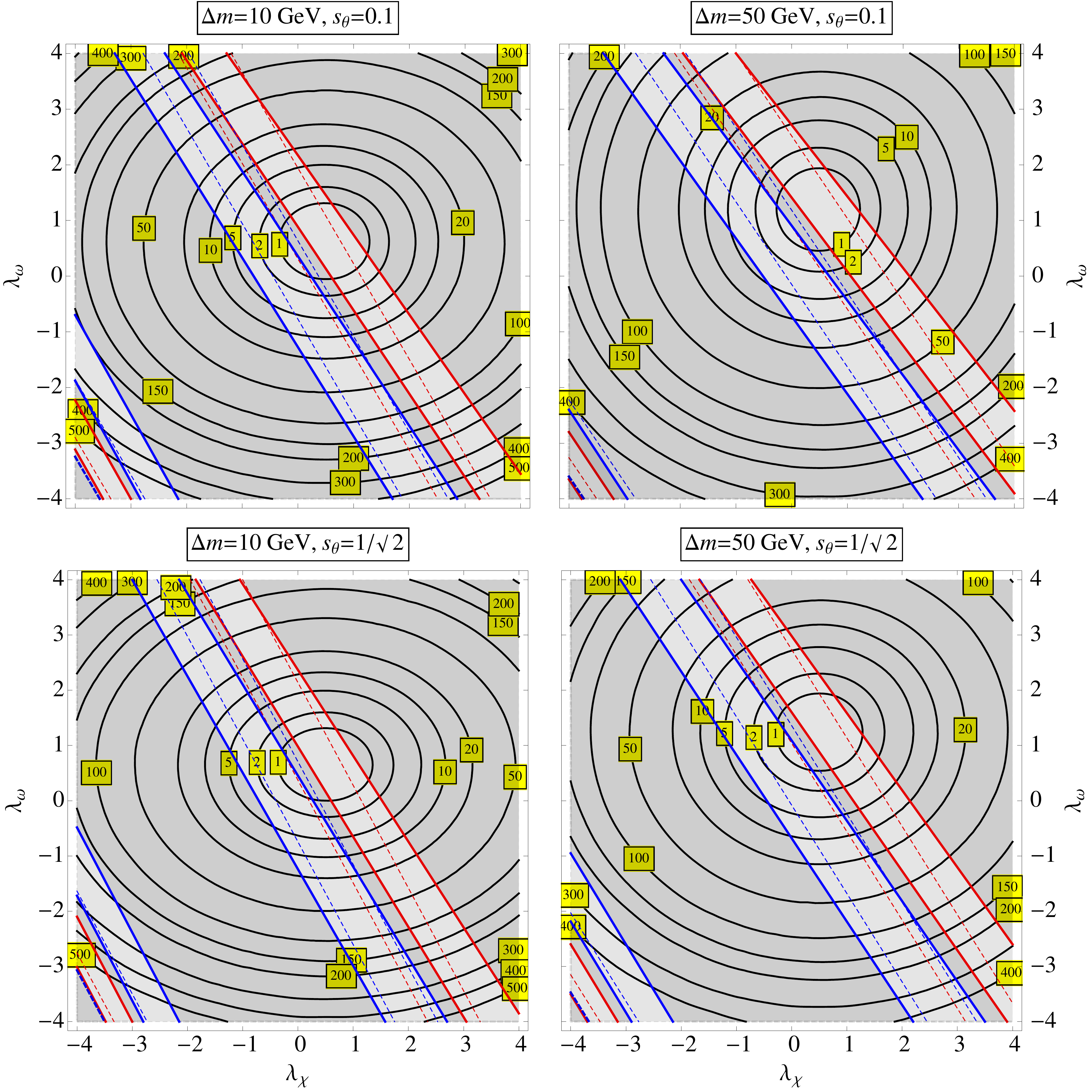}
 \caption{Scanned contour plot in $\lambda_\omega$--$\lambda_\chi$ for  the ratio Higgs pair productions due LQs and the SM. Here the mass of the lightest LQ is chosen to be 200~GeV.
 process.}\label{fig:gghhscan}
\end{figure}

The leading order (LO) partonic amplitude for Higgs productions cross-section and the diphoton decay rates are given by:
\begin{eqnarray}\label{eq:sigma1}
\sigma_{gg\rightarrow h}&=&\frac{G_F\alpha_s^2}{126\sqrt{2}\pi}\left|\frac{1}{2}A_{\frac{1}{2}}\left(x_t\right)+\sum^{}_{i} C_i\frac{\lambda_iv^2}{4m_{s_i}^2}A_0\left(x_{s_i}\right)\right|^2,\\
\Gamma_{\gamma\gamma}&=&\frac{G_F\alpha^2m_h^3}{126\sqrt{2}\pi^3}\left|A_{1}\left(x_W\right)+\frac{4}{3}A_{\frac{1}{2}}\left(x_t\right)+\sum^{}_{i} \frac{\lambda_i}{g_w}\frac{m_W^2}{m_{s_i}^2}d_iQ_i^2A_0\left(x_{s_i}\right)\right|^2.
\end{eqnarray}
where $x_\phi=4m^2_\phi/m^2_h$ for $\phi=t,s_i,W$ and the well known loop functions are listed  $A_{(0,\frac{1}{2},1)}$ in the Appendix. The NLO and NNLO corrections are substantial leading to an enhancement of $K\sim2$~\cite{Dawson:1998py, Djouadi:1999rca, Dittmaier:2011ti, Branco:2011iw, Baglio:2012np, Dolan:2012rv, Papaefstathiou:2012qe, Goertz:2013kp, deFlorian:2013uza, Grigo:2013rya, Barger:2013jfa}. Since we are primarily interested in the changes from the additional states we take the ratio of the new rates compared and that of the SM where the NLO and NNLO corrections are expected to largely cancel out.  
The values of the loop functions for $W$ and top are $A_1(x_W)=-8.3$ and $A_{1/2}=1.38$.

\begin{figure}[!ht]
 \centering
 \includegraphics[height=20cm]{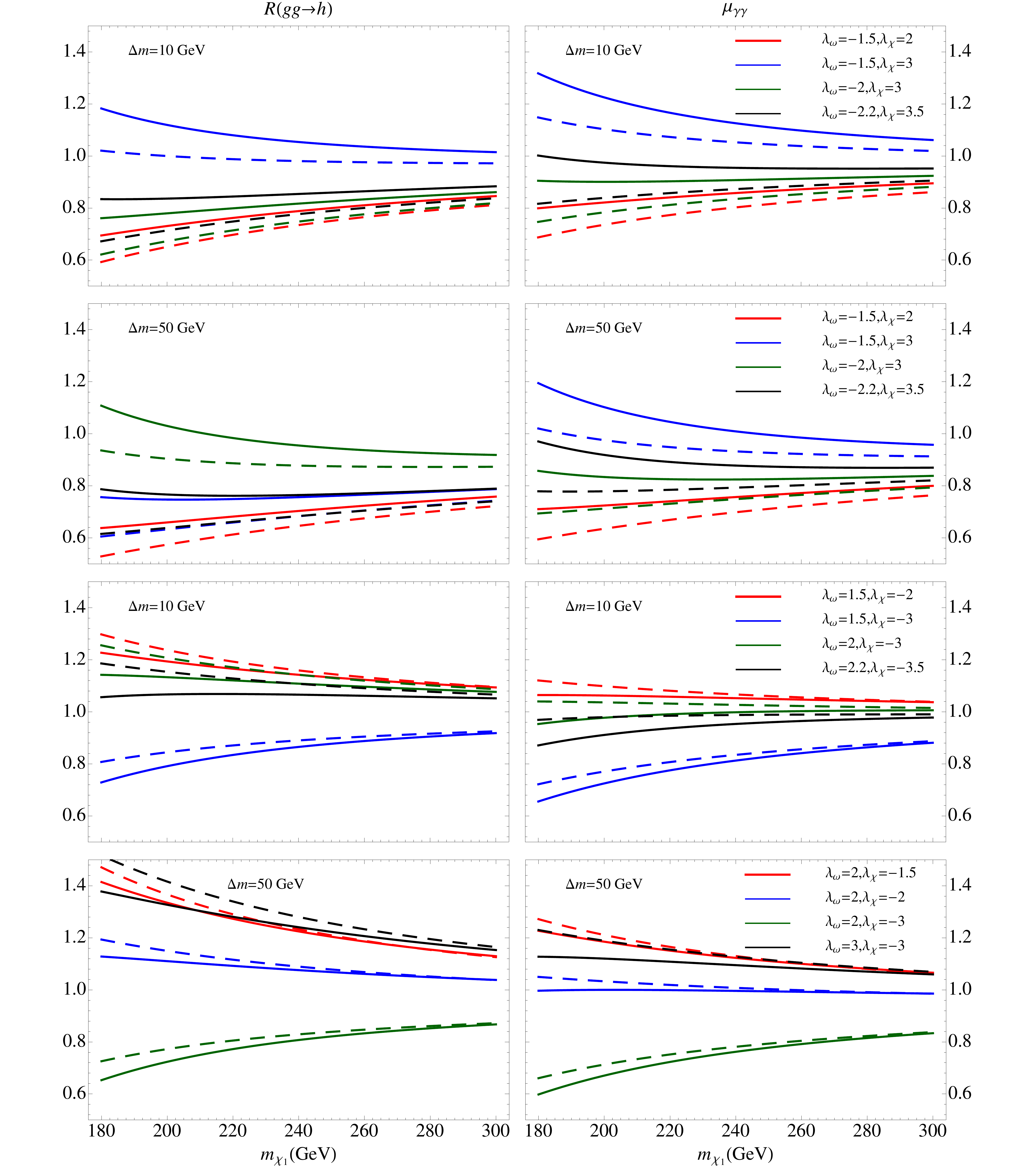}
 \caption{The Higgs production rate  and its significance in the presence of several LQs compared to the SM.
 process. The solid (dashed) curves are for the LQ mixing angle with $\sin\theta_\chi=0.1(1/\sqrt{2}).$}
 \label{fig:ggh}
\end{figure}

In the Standard Model contributions from the top quark triangle and box diagrams largely cancel  each other for $\sim$125~GeV Higgs mass resulting in a few fb production cross section. It is estimated that with few thousand~fb$^{-1}$ at $14$~TeV, a 3$\sigma$ evidence may be reached~\cite{Dawson:1998py, Djouadi:1999rca, Dittmaier:2011ti, Branco:2011iw, Baglio:2012np, Dolan:2012rv, Papaefstathiou:2012qe, Goertz:2013kp, deFlorian:2013uza, Grigo:2013rya, Barger:2013jfa}. This situation may be altered by additional colored particles. The parton level cross-section is given by  
\begin{eqnarray}\label{eq:sigma2}
\frac{{d\hat\sigma}_{gg\rightarrow hh}}{d\hat t}=\frac{G_F^2\alpha_s^2}{256(2\pi)^3}\left(\left|\frac{3m^2_h}{\hat{s}-m^2_h+im_h\Gamma_h}F_{tri}+F_{box}\right|^2+\left|G_{box}\right|^2\right)
\end{eqnarray}
There are two types of amplitudes, $F$ and $G$, corresponding to the same and opposite polarization of the incoming gluons respectively. The same polarization part comes from triangular and box diagrams while the opposite one does only from box diagrams. Here the triangular is meant to be the one with the Higgs propagator and therefore is proportional to the Higgs self coupling. The other triangle diagrams not proportional to the Higgs self coupling are combined with the box diagrams. The amplitudes in the SM and in models with additional colored scalars are given in the Appendix. 

For the masses we take hierarchy $m_\omega>m_{\chi_2}>m_{\chi_1}$. In addition I choose $\Delta m\equiv m_\omega-m_{\chi_2}=10$ and $50$~GeV for small and large splitting and a  constant value of 10~GeV for the mass splitting between the lighter two $m_{\chi_2}-m_{\chi_1}=10$~GeV. I take two different values for the LQ mixing $\sin\vartheta=0.1$ and $1/\sqrt{2}$ for small and large mixings respectively.

Previous studies have considered an effect of a single colored particles, where one is forced to have a specific couplings not to upset the Higgs production rate. For example, the new physics contribution is chosen to be roughly twice larger and opposite in sign to have unaltered rate. This inevitably affects diphoton channel. In particular among possible color scalars only octet candidate was a good choice~\cite{Kribs:2012kz}.  For these models, stability of vacuum requires increasingly stronger portal couplings as the mass is increased~\cite{Reece:2012gi}. This is because one needs to keep the new contribution to the Higgs production more or less constant for higher mass values which is possible only if the corresponding portal coupling is simultaneously increased. This is not required in our case, since we have several new contributions which can be kept under control by a judicious choices of the various portal couplings as far as the Higgs production and diphoton channels are concerned. 

We first scan over the $\lambda_\omega$ and $ \lambda_\chi$ parameter space for the Higgs pair production and super impose the allowed regions by both CMS and ATLAS experiments by the diphoton and $ZZ^*$ channels. The result is show in Figure~\ref{fig:gghhscan}. The lightest LQ mass is chosen to be $m_{\chi_1}=200$~GeV. The parameter scan has been done using MadGraph~5~\cite{Alwall:2011uj} with CTEQ6L1 PDF set~\cite{Sjostrand:2006za}. The Madgraph implementation of the Higgs pair production in the SM has been modified to include contributions from the LQ. The code has been checked against previously known results such as in Ref~\cite{Kribs:2012kz} and was found to be in an excellent agreement.

As we can see there are regions in the parameter space where the single Higgs production and decay rates are compatible with either of CMS and ATLAS experiments. Depending on the values for the couplings the Higgs pair production may become substantially enhanced. The shape of the regions are easily understood. The single Higgs production rate and decay to diphoton and $ZZ^*$ channels will be affected less if the contributions from the LQs largely cancel each other. This fact is reflected in the stripe regions. There is another possibility that the total LQ contribution is twice bigger than the SM amplitude and but opposite in sign as has been done in Ref~\cite{Kribs:2012kz, Dorsner:2012pp}. This possibility is represented by the allowed region in the lower right corner of the scanned plots in Figure~\ref{fig:gghhscan}  where both $\lambda_\omega$ and $ \lambda_\chi$ are large and negative. Since this region will be pushed to higher values as the LQ mass is increased we do not consider this region further and concentrate on the stripe regions. 

While these regions obviously should become larger for heavier choice of the LQ mass $m_{\chi_1}$, to make sure that the allowed parameters from the scanning are not accidental for the particular choice I made, the single Higgs production  is calculated for several set of $\lambda_\omega$ and $\lambda_\chi$ with $m_{\chi_1}$ changing from $180$ to $300$~GeV.  The results are plotted and shown in Figure~\ref{fig:ggh}. The plots in the right column labeled as $R(gg\rightarrow h)$ are the single Higgs production rate and the plots in the left column are for the corresponding signal significance in the diphoton channel labeled as $\mu_{\gamma\gamma}$ both compared to the SM. As we see the rates are within the one $\sigma$ range of the either of the two experiments at the LHC and approach to the SM values with increasing mass as one would expect.

\begin{figure}[t!]
 \centering
 \includegraphics[height=9cm]{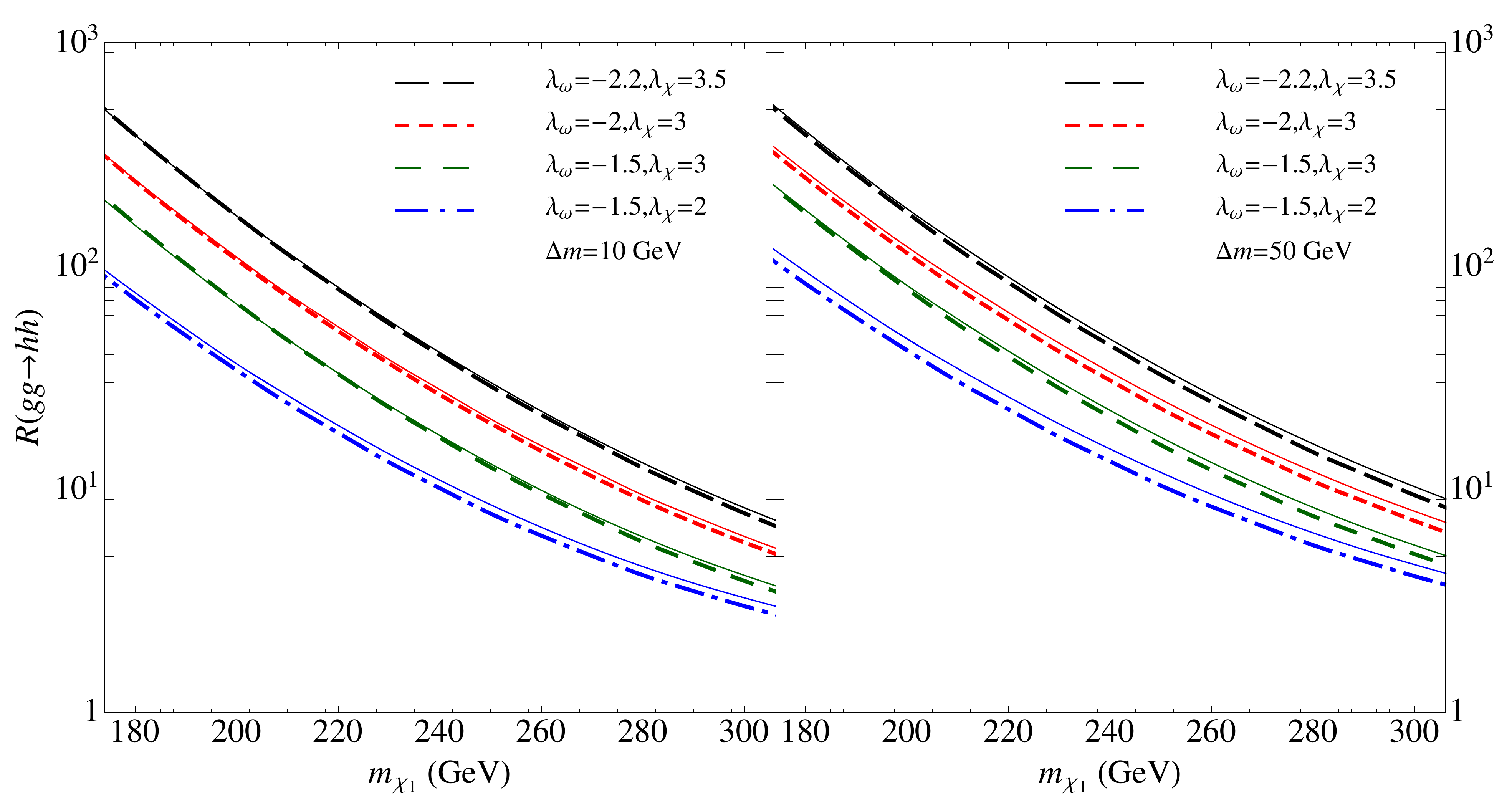}
 \caption{The ratio of Higgs productions due LQs and the SM for negative $\lambda_\omega$ and positive $\lambda_\chi$ for several different choices for the mixing parameter $s_\theta=0.1$. Thin lines with the same colors to the various dashed lines are obtained when the maximal mixing $s_\theta=1/\sqrt{2}$ is chosen. }
 \label{fig:gghh1}
\end{figure}

\begin{figure}
 \centering
 \includegraphics[height=9cm]{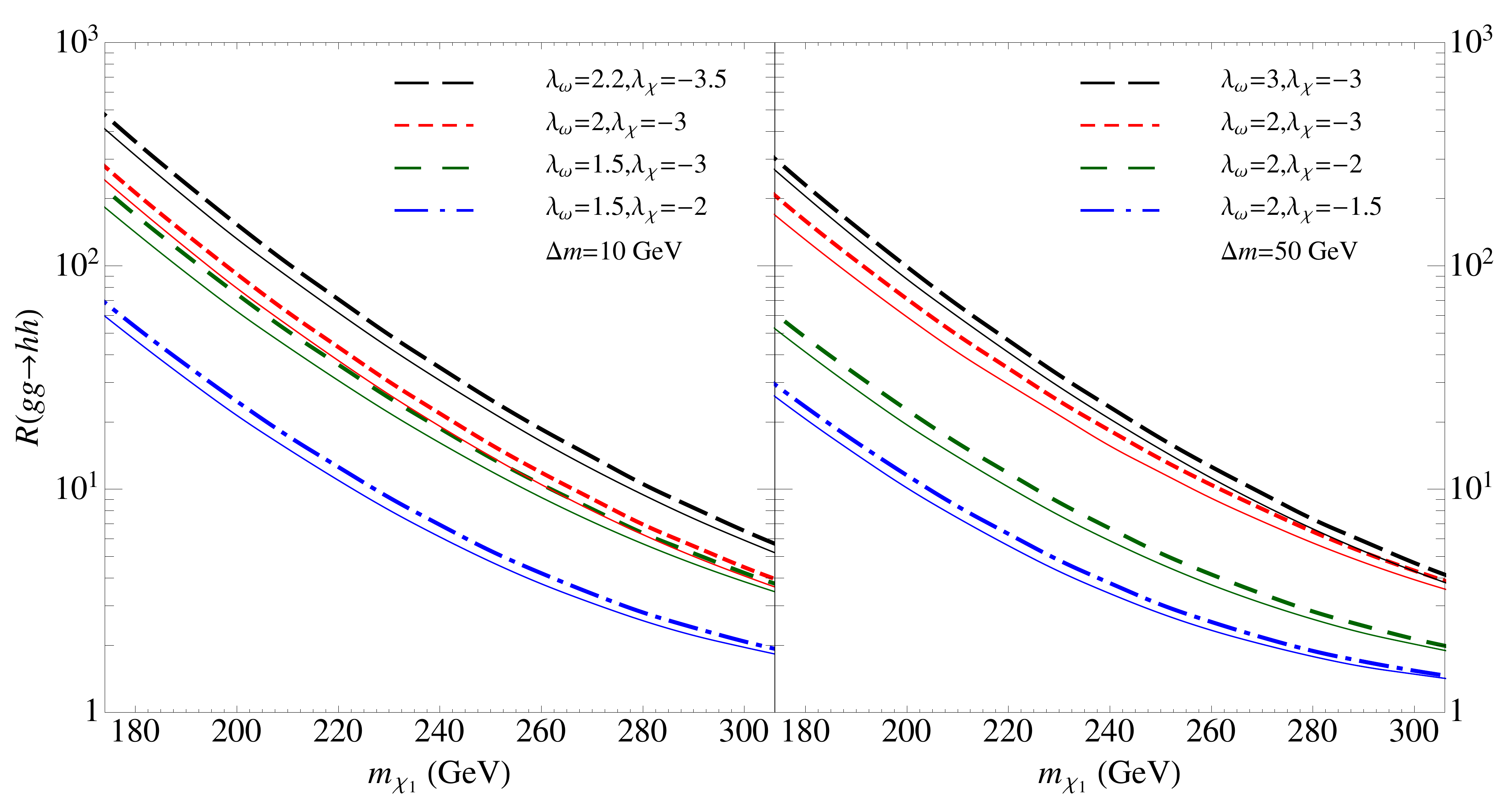}
 \caption{The ratio of Higgs pair productions due LQs and the SM for positive $\lambda_\omega$ and negative $\lambda_\chi$.
 process.}
 \label{fig:gghh2}
\end{figure}

Next, I estimate the Higgs pair production for the same set of parameters. The results are shown in Figures~\ref{fig:gghh1} and \ref{fig:gghh2}. These are the main results of the present work. As we can see the rate may be enhanced quite substantially compared to the SM expectation even the single Higgs production is affected moderately. The cancellation due to the opposite sign for  $\lambda_\omega$ and $\lambda_\chi$, which kept the single Higgs rate largely unchanged, is still operational for the triangular loop diagram contributions to the pair production. However, there are diagrams quadratic in the portal couplings whenever the final state Higgses come from different vertices. They will contribute constructively even if the single Higgs production remain the same as in the SM. The largest values I chose for the portal couplings require even larger value for the quartic couplings for LQ to make the vacuum at least metastable~\cite{Reece:2012gi} since we have a negative portal coupling. If we generously allow and take values up to $4\pi$ for the quartic couplings the metastability of the vacuum is guaranteed.


A detailed signal simulation for the LQ pair productions for the LHC experiments is beyond the scope of the present paper. Nevertheless, few comments are in order. The search for $pp\to LQ\overline  LQ\rightarrow t\tau^- \bar t \tau^+$  signal has not been done by either of the two collaborations. The pair production cross--section is roughly an order of magnitude below that of $t\bar t$ if $m_{LQ}$ little above $m_t$. Then the signal is hard to distinguished  from $t\bar t$ as the taus would not be energetic enough.  Therefore such light LQs are still a possibility. For higher values starting around 200~GeV and upto $\sim$260~GeV, recently performed searches for the Higgs production in association with a top pair $gg\to t\bar t h$, with Higgs decaying to tau pair~\cite{cmsttH}, may rule out some mass regions whenever the chosen cuts are applicable to the event generated by the LQ pairs. This process has the same final state as the pair produced leptoquarks decaying to $t\tau$. The observed upper bound is $\sigma/\sigma_{SM}=13$ with signal strength $\mu=-0.7\substack{+6.2 \\ -5.3}$. Therefore, taking $\sigma(pp\to t\bar t h)\simeq80$~fb at LO and $BR(h\to \tau\tau)\simeq7$~\%, one may conclude that $\sigma(pp\to LQ\overline LQ)$ should not exceed a few hundred~fb to $O(1)$~pb at most. The exact constraint and implication of this process needs a thorough analysis and I do not attempt such study in this paper.

\section{Conclusions and Discussions}
\noindent
The discovery of a scalar particle by CMS and ATLAS experiments at the LHC which appears to be essentially consistent with the SM Higgs picture within experimental margin of error is a triumph in our understanding of the fundamental dynamics. Undoubtedly, more precise measurements of the various production and decay channels are needed to nail down it as the Higgs of the SM. On the other hand, the confirmation itself still leave many questions unanswered which can be addressed with new dynamics or particles at the TEV range.  

Colored particles around TeV scale have been studied in context of different theories for various reasons. With current 7 of 5~fb$^{-1}$ and 8 TeV of 19.4~fb$^{-1}$ data, these can be probed if they are not too heavy. Among these the colored particles interacting with the SM Higgs doublet may cause an enhancement for Higgs pair production.

In the present paper, I have considered a several  scalar LQs in which their portal couplings are such that its effect on  the single Higgs production is within the limits given by the either CMS or ATLAS experiment. Even in this case it has been found that the Higgs pair production can be modified substantially.
For several set of values for the portal couplings it has been shown that the rate may reach one to two orders magnitude higher than what it is in the SM. The two portal couplings are chosen to have an opposite sign which give reasonable single Higgs production rate via gluon fusion.

These are done via the following procedure. Upon scanning over these couplings for a low mass value the allowed regions by the Higgs porduction and decay to diphoton and $ZZ^*$ are obtained. Several set of values are chosen from these regions. We ignore the possibility of having both portal couplings are negative such that it produces a contribution twice as big as the SM one but with  opposite sign. For the chosen values for the couplings the single Higgs productions have been plotted for masses upto 300~GeV where the rates remain within the experimentally allowed region. Once this established, the Higgs pair production has been studied. For all the values the rates have been found to be enhanced by various values. For the sets with larger values, it may reach two orders of magnitude at lower range of LQ masses with moderate effect on the single Higgs production.

The effect becomes negligible above around the mass of 300~GeV. For this value, the enhancements range from few to at most an order of magnitude. In this case we have to wait the 14~TeV run of the LHC experiments and high luminosity. Then the LQ will be ruled out or discovered before we reach the Higgs pair production discovery.

The present work demonstrates that the light colored particles with large portal couplings may reveal additional dynamics in the scalar potential. These are interesting due to their potential role in EWSB itself or in the thermal phase transition in the early universe. The model considered here is an example. From this study, one can see that any models with several color colored particles with strong couplings to Higgs can have sustantial effect on the Higgs pair production.

\section{Appendix}

\noindent
Here we collect the formulae we used in our numerical calculations for the single and pair Higgs productions. The loop functions in Eq.~(\ref{eq:sigma1}) for the single Higgs productions are give by:
\begin{eqnarray}\label{eq:Afunct1}
A_1(x)&=&-\left(2+3x+3x(2-x)f(x)\right),\\
A_{1/2}&=&2x\left(1+(1-x)f(x)\right),\\
A_0&=&-x\left(1-xf(x)\right),\\
f(x)&=&
\begin{cases}
   \,\,\,\,\,\,\,\,\,\,\,\arcsin^2\left(1/\sqrt{x}\right),& \text{if } x\geq 1\\
    -\dfrac{1}{4}\left(\log\dfrac{1+\sqrt{1-x}}{1-\sqrt{1-x}}-i\pi\right)^2,    & \text{if } x < 1
\end{cases}
\end{eqnarray}

The Higgs pair production amplitudes are separated into two parts $F$ and $G$ from the same and opposite initial gluon polarizations respectively.  
The contributions from the SM for the process $g(p_A)g(p_B)\rightarrow h(p_C)h(p_D)$ are given by:
\begin{eqnarray}\label{eq:SM2h}
F_{tri}&=&\frac{2m_t^2}{s}\left(2+\left(4m_t^2-{s}\right)C_{AB}\right),\\
F_{box}&=&\frac{2m_t^2}{s}\left(2+4m_t^2C_{AB}
     -\left(s+2m_h^2-8m_t^2\right)m_t^2\left(D_{ABC}+D_{BAC}+D_{ACB}\right)
     \right. \nonumber\\ 
     &+& \left. \frac{m_h^2-4m_t^2}{s}\left(\left(t-m_h^2\right)\left(C_{AC}
     +C_{BD}\right)+\left(u-m_h^2\right)\left(C_{BC}+C_{AD}\right)\right.\right.\nonumber\\
    & -&\left.\left.  \left(tu-m_h^4\right)D_{ACB}\right)\right)\\
    G_{box}&=& \frac{m_t^4}{s(tu-m_h^4)}\left(
          \frac{(t^2+m_h^4-8tm_t^2)}{m_t^2}
          (sC_{AB}+(t-m_h^2)(C_{AC}+C_{BD})-s t D_{BAC})\right. \nonumber\\ 
     &+& \left.\frac{(u^2+m_h^4-8um_t^2)}{m_t^2}
     	(sC_{AB}+(u-m_h^2)(C_{BC}+C_{AD})-s u D_{ABC})\right. \nonumber\\ 
     &-& \left.
        \frac{(t^2+u^2-2m_h^4)(t+u-8m_t^2)}{m_t^2}C_{CD}\right. \nonumber\\ 
     &-& \left.
        2(t+u-8m_t^2)(t u-m_h^4)\left(D_{ABC}+D_{BAC}+D_{ACB}\right)\right)
\end{eqnarray}
Additional colored scalar particles contribute the following amplitudes:
\begin{eqnarray}\label{eq:LQ2h}
F^S_{tri}&=&-\frac{\lambda_SC_sv^2}{m_S^2}(2m_S^2C_{AB}+1),\\
F^S_{box}&=&-\frac{\lambda_SC_sv^2}{m_S^2}(2m_S^2C_{AB}+1)
     -\frac{2C_s(\lambda_Sv^2)^2}{s}\left(
     m_S^2\left(D_{ABC}+D_{BAC}+D_{ACB}\right)\right. \nonumber\\ 
           & -& \left. \frac{t-m_h^2}{s}C_{AC}-\frac{u-m_h^2}{s}C_{BC} + \frac{u t-m_h^4}{2s}D_{ACB}\right),\\
             G^S_{box}&=& -\frac{2C_s(\lambda_Sv^2)^2}{s}(
      m_S^2\left(D_{ABC}+D_{BAC}+D_{ACB}\right)-C_{CD}\nonumber\\
     & +&\frac{1}{2(tu-m_h^4)} (s t^2D_{BAC}+s u^2D_{ABC}\nonumber\\
     & +&s(s-2m_h^2)C_{AB}+s(s-4m_h^2)C_{CD}\nonumber\\
     & -&2t(t-m_h^2)C_{AC}-2u(u-m_h^2)C_{BC})    )
\end{eqnarray}
Here $C_{AB}$ and $D_{ABC}$ etc are Passarino-Veltman 3 and 4--point functions and are given by
\begin{eqnarray}\label{eq:PaVe}
C_{AB}&\equiv&\int \frac{d^4q}{i\pi}\frac{1}{(q^2-m^2)((q+p_A)^2-m^2)((q+p_A+p_B)^2-m^2)},\\
D_{ABC}&\equiv&\int \frac{d^4q}{i\pi}\frac{1}{(q^2-m^2)((q+p_A)^2-m^2)((q+p_A+p_B)^2-m^2)}\nonumber \\
&\times &\frac{1}{((q+p_A+p_B+p_C)^2-m^2)}.
\end{eqnarray}
Here $m=m_t$ and $m_S$ substitutions should be used for the top quark and colored scalar contribution respectively.

\section*{Acknowledgment}
The author would like to thank Prof. Kai-Feng Chen for numerous discussions regarding the search of LQs at LHC and the authors of Ref~\cite{Papaefstathiou:2012qe} for providing their MadGraph implementation of the Higgs pair production for the SM and explaining the details. The travel support from The Mongolian Fund for Science and Technology and the local support from the organizers of the PASCOS 2013, where the work is presented, are acknowledged. The work was supported in part by the National Science Council
of Taiwan under Grant No.~NSC 101-2811-M-002-028.



\end{document}